\documentclass[pra,aps,showpacs,twocolumn,floatfix]{revtex4}
\usepackage{graphicx}
\usepackage[ansinew]{inputenc}
\usepackage{array}
\usepackage{color}
\usepackage{amsmath}
\usepackage{amsxtra}
\usepackage{amstext}
\usepackage{amssymb}
\usepackage{latexsym}
\usepackage{dsfont}
\usepackage{verbatim}
\usepackage{comment}

\begin{document}

\title
{Spin Squeezing with Coherent Light via Entanglement Swapping}

\author{M.E. Ta\c{s}g\i n and P. Meystre}
\affiliation{B2 Institute, Department of Physics, and College of Optical Sciences\\The University of Arizona, Tucson, Arizona 85721, USA}

\date{\today}

\begin{abstract}
We analyze theoretically a scheme that produces spin squeezing via the continuous swapping of atom-photon entanglement into atom-atom entanglement, and propose an explicit experimental system where the necessary atom-field coupling can be realized. This scheme is found to be robust against perturbations due to other atom-field coupling channels. 
\end{abstract}


\maketitle

\section{Introduction}
A spin ensemble prepared in an atomic coherent state (ACS) \cite{mandelwolf} can be used to perform measurements with precision limited by the standard quantum limit (SQL), $\langle\Delta S_\perp\rangle \geq \left|\langle\mathbf{S}\rangle\right|/2$, where $\mathbf{S}$ is the spin vector and $S_\perp$ is the orthogonal spin component that is measured. One way to overcome that limit is by using squeezed spin states (SSSs) \cite{ueda-pra-1993} in which the uncertainty in one of the orthogonal spin components is reduced below the SQL. It has however proven difficult to achieve spin squeezing experimentally.  Its realization requires some kind of nonlinear coupling between spins,~\cite{ueda-pra-1993,sorensen-nature-2001} but intrinsic spin-spin interactions are normally quite weak, resulting typically in small amounts of squeezing \cite{sorensen-nature-2001,zoller-pra-2003,esteve-nature-2008,duan-prl-2000-bec}. 

This difficulty can be circumvented by exploiting the stronger spin-spin couplings that can be mediated by optical interactions. This was recognized early on, and the generation of SSS by quantum state transfer from squeezed light has been studied since the early 1990s~\cite{wineland-pra-1992,grom-1995,kuzmich-prl-1997,hald-prl-1999}. Recent work also considered both theoretically~\cite{takahasi-pra-1999,kuzmich-epl-1998} and experimentally~ \cite{kuzmich-prl-2000,vuletic-prl-2010-QND,PS1} the collapse of the state of a spin system to a SSS resulting from a measurement on the light field coupled to that ensemble. 

A promising method proposed by Takeuchi~\textit{et al.}~\cite{takeuchi-prl-2005} involves interacting an optical field twice with the atomic ensemble, the second interaction taking place after the optical field polarization is rotated and it is reflected by a polarizer/mirror combination. The first interaction entangles a photon with an atom in the ensemble, and the second interaction couples that same photon to a second atom. As a result the photon swaps the entanglement \cite{ekert-prl-1993,bennett-prl-1993,knight-pra-1998,zeilinger-prl-1998,mustecapPRA2009} between the two atoms and produces a SSS. This method was recently demonstrated in a cavity configuration by Vuletic and coworkers~\cite{vuletic-prl-2010}.

In this note we consider an alternative scheme that couples the optical field and the atoms in such a way that entanglement swapping takes place in a {\textit single path}. Like the scheme of Ref.~ \cite{takeuchi-prl-2005}, this method requires only a coherent light pulse and linear optics elements, but it is simpler than Takeuchi's approach in that does not require the mirror and polarizer, nor does it require an optical resonator as in the experiments of Vuletic {\it et al.} As such, it should be widely applicable. The required interaction can be realized for example in alkali atoms for an appropriate choice of atom-field detunings. 

The interaction Hamiltonian that achieves that goal has the form
\begin{equation}
\hat{H}=\alpha \left(\hat{J}_+\hat{S}_+ + \hat{J}_-\hat{S}_-\right),
\label{eq:hamiltonian}
\end{equation}
where $\alpha$ is the coupling strength, $\hat{S}_{\pm}$ are the ladder operators for the spin-$S$ system, and
\begin{eqnarray}
\label{schwinger}
\hat{J}_+&=&\hat{J}_-^{\dagger}=\hat{a}_-^{\dagger}\hat{a}_+ \nonumber \\
\hat{J}_z&=&\frac12(\hat{a}_-^{\dagger}\hat{a}_- - \hat{a}_+^{\dagger}\hat{a}_+)
\end{eqnarray}
are similarly the Schwinger representation operators for the two optical modes of polarizations  $\sigma_+$ and $\sigma_-$,  where $\hat{a}_+ , \hat{a}_+^{\dagger}$ and $\hat{a}_- , \hat{a}_-^{\dagger}$ are the corresponding  annihilation and creation operators. The operators $\hat{\mathbf{S}}$ and $\hat{\mathbf{J}}$ obey angular momentum commutation relations 
\begin{eqnarray}
\left [\hat{S}_i,\hat{S}_j\right ] &=& i\epsilon_{ijk}\hat{S}_k, \nonumber \\
\left [\hat{J}_i,\hat{J}_j\right ] &=& i\epsilon_{ijk}\hat{J}_k
\end{eqnarray}  
where the indices $\{i,j\}$ stand for the $x,y,$ and $z$ vector components.

This paper is organized as follows: We begin in section II by discussing a possible experimental realization of the Hamiltonian~(\ref{eq:hamiltonian}) involving electric dipole transitions in alkali atoms such as $^{87}$Rb. Section III summarizes some key aspects of the average spin dynamics, and compares it to the situation for the model system ${\hat H} \propto {\hat S}_z^2$ originally considered by Kitagawa and Ueda \cite{ueda-pra-1993}. Section IV then turns to a discussion of the generation of spin squeezing per say, and comments on its physical origin in the swapping of entanglement from the spin-photon system to pairs of atoms. Finally, Section V is a summary and conclusion.

\section{Experimental realization}

\begin{figure}
\includegraphics[width=3.2in]{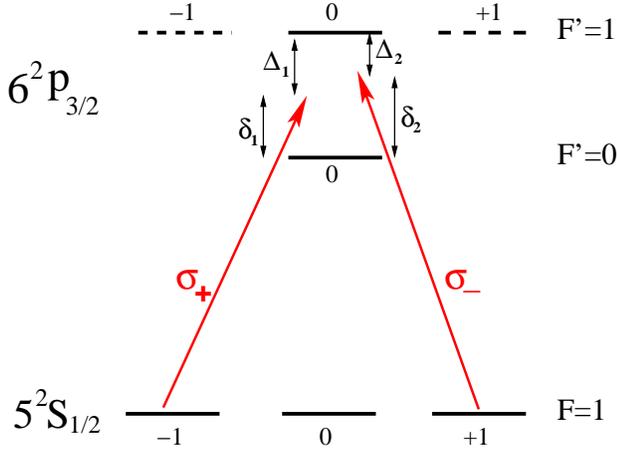}
\caption{Hyperfine states of ${}^{87}\text{Rb}$ on the $\text{D}_2$ transition line. The two ground states $|F=1,m_F=\pm 1\rangle$ are optically coupled to the excited states $|e_1\rangle=|F'=0,m_{F'}=0\rangle$ and $|e_2\rangle=|F'=1,m_{F'}=0\rangle$ by $\sigma_\pm$-polarized light. The detunings of the two modes are chosen such that the diagonal terms in Eq.~(4) vanishes, resulting in the model Hamiltonian~(\ref{eq:hamiltonian}). }
\label{fig1}
\end{figure}

One possible way to realize an interaction Hamiltonian of the form (\ref{eq:hamiltonian}) is by coupling the two $5^2S_{1/2}$ hyperfine states $|F=1,m_F=\pm 1\rangle$ of $^{87}$Rb to the  $F'=0$ and $F'=1$ hyperfine manifolds of the  $6^2P_{3/2}$ state with two optical fields of opposite circular polarizations and detunings $-\Delta_{1,2}$ and $\delta_{1,2}$, see Fig.~ \ref{fig1}. The electric dipole coupling constants between the $|F=1, m_F=-1\rangle$ ground state and the  $|e_1\rangle=|F'=0,m_{F'}=0\rangle$ excited state, and between the $|F=1, m_F=-1\rangle$ ground state  and the $|e_2\rangle=|F'=1,m_{F'}=0\rangle$ excited state, are $g_{-,e_1}=\sqrt{1/6}$ and $g_{-,e_2}=\sqrt{5/24}$, respectively. For the transitions from the $m_F=1$ ground state, the corresponding coupling constants are $g_{+,e_1}=\sqrt{1/6}$ and $g_{+,e_2}=-\sqrt{5/24}$~\cite{danielsteckRb87}.

Since the magnitudes of the ratio of the couplings are equal for both $\sigma_\pm$ transitions, the situation simplifies considerably for the choice of detunings $\Delta_1=\Delta_2=\Delta$ and $\delta_1=\delta_2=\delta$. After adiabatic elimination of the excited levels the atom-field coupling Hamiltonian becomes then ($\hbar = 1$)
\begin{eqnarray}
\hat{H}&=&\left( \frac{\left |g_{-,e_1}\right |^2}{\delta} - \frac{\left |g_{-,e_2}\right|^2}{\Delta} \right) \hat{\psi}_{-}^{\dagger}\hat{\psi}_{-}\hat{a}_+^{\dagger}\hat{a}_+  \nonumber \\
&+& \left( \frac{\left |g_{+,e_1}\right |^2}{\delta} - \frac{\left |g_{+,e_2}\right |^2}{\Delta} \right) \hat{\psi}_{+}^{\dagger}\hat{\psi}_{+}\hat{a}_-^{\dagger}\hat{a}_-  \\
&+&\left( \frac{g_{-,e_1}^*g_{+,e_1}}{\delta} - \frac{g_{-1,e_2}^*g_{+,e_2}}{\Delta} \right) \hat{\psi}_{-}^{\dagger}\hat{\psi}_{+}\hat{a}_-^{\dagger}\hat{a}_+  +{\rm h.c.} \nonumber
\label{eq:couplingH}
\end{eqnarray}
where $\hat{\psi}_\pm$ and $\hat{\psi}_\pm^{\dagger}$ are annihilation and creation operators for the atomic hyperfine states $5{}^2S_{1/2}$ $|F=1,m_F=\pm1\rangle$. A further simplification follows from the fact that the diagonal terms in that Hamiltonian vanish for 
\begin{equation}
\frac{\delta}{\Delta}=\frac{\left |g_{\pm,e_1}\right |^2}{\left |g_{\pm,e_2}\right |^2}=20,
\end{equation}
with the off-diagonal terms remaining non-zero since $g_{-,e_1}^*g_{+,e_1}$ and $g_{-,e_2}^*g_{+,e_2}$ have opposite signs~\cite{PS3,PS4}. Under these conditions, the Hamiltonian~(\ref{eq:couplingH}) maps precisely to the model Hamiltonian (\ref{eq:hamiltonian}), provided that it is generalized to the case of $N$ identical atoms and that we introduce the Schwinger representation (\ref{schwinger}).

\section{Spin dynamics}

The Hamiltonian (\ref{eq:hamiltonian}) is not solvable exactly, in contrast to the more widely studied $\hat{J}_z\hat{S}_z$ interaction, and consequently our discussion is largely restricted to the presentation of selected numerical results. We start with the expectation value of the spin operators $\langle S_x\rangle$, $\langle S_y\rangle$ and $\langle S_z\rangle$, concentrating on features that will prove useful in the understanding of squeezing in the following section. We also compare these values to the corresponding results for the Hamiltonian ($\hbar = 1$)
\begin{equation}
{\hat H}_{ku} = \alpha {\hat S}_z^2
\label{HKU}
\end{equation}
which has been discussed in detail by Kitagawa and Ueda in Ref.~\cite{ueda-pra-1993} and for which the squeezing features are well understood.

\begin{figure}
\includegraphics[width=3.6in]{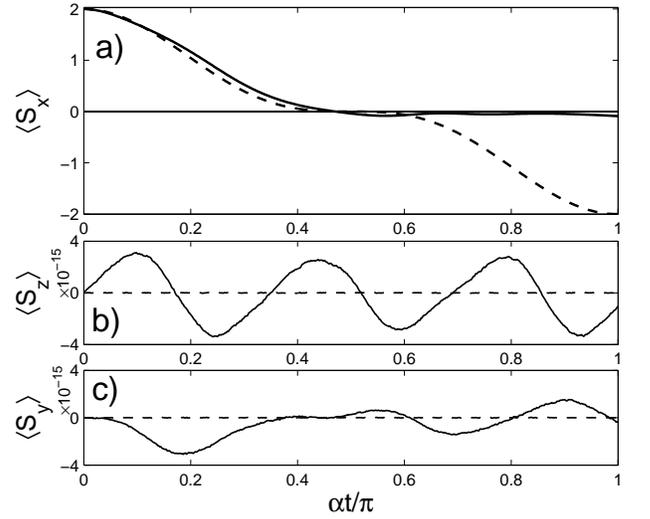}
\caption{Time evolution of the expectation values $\langle S_x \rangle$, $\langle S_y \rangle$ and $\langle S_z \rangle$ for the Hamiltonians~(\ref{eq:hamiltonian}), solid line, and (\ref{HKU}), dashed line, for $S=2$ and $J=2$. Note the vastly different vertical scales on the plot for $\langle S_x \rangle$, compared to those for $\langle S_y \rangle$ and $\langle S_z \rangle$, whose amplitudes are 15 orders of magnitude smaller. Time in units of $\pi/\alpha$.}
\label{fig2}
\end{figure}

\begin{figure}
\includegraphics[width=3.6in]{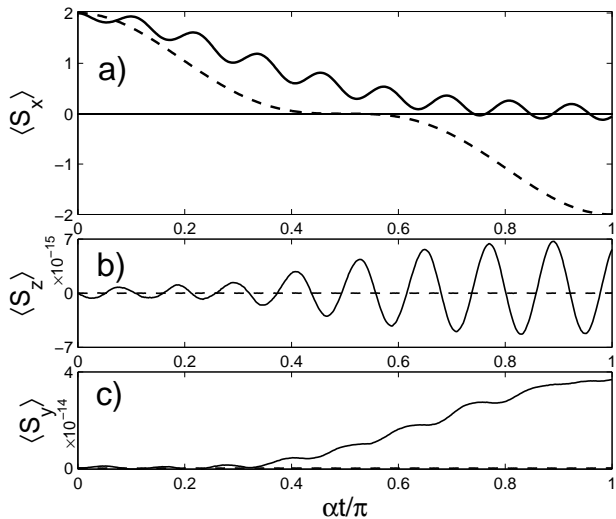}
\caption{Time evolution of the expectation values $\langle S_x \rangle$, $\langle S_y \rangle$ and $\langle S_z \rangle$ for the Hamiltonians~(\ref{eq:hamiltonian}), solid line, and (\ref{HKU}), dashed line, for $S=2$ and $J=8$. Increasing $J$ induces an oscillatory behavior in $\langle S_x \rangle$. Note again the vastly different vertical scales for the three plots. Time in units of $\pi/\alpha$.}
\label{fig3}
\end{figure}

\begin{figure}
\includegraphics[width=3.6in]{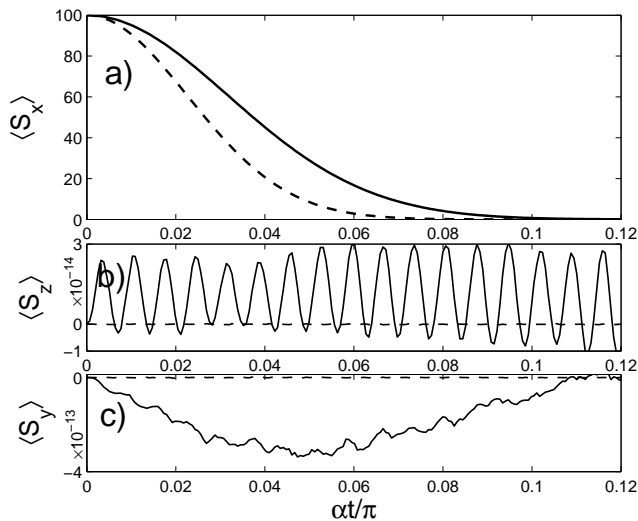}
\caption{ Time evolution of the expectation values $\langle S_x \rangle$, $\langle S_y \rangle$ and $\langle S_z \rangle$ for the Hamiltonians~(\ref{eq:hamiltonian}), solid line, and (\ref{HKU}), dashed line, for $S=100$ and $J=100$. Note again the vastly different vertical scales for the three plots. Time is in units of $\pi/\alpha$.}
\label{fig4}
\end{figure}

To set the stage for the discussion, Figs.~ \ref{fig2}a and \ref{fig2}b,c compare the expectation values $\langle S_x \rangle$, $\langle S_y \rangle$ and $\langle S_z \rangle$ for the Hamiltonians~(\ref{eq:hamiltonian}) and (\ref{HKU}) for the small spin value $S=2$ and $J=2$. Figure~\ref{fig3} shows these same expectation values for $S=2$ and $J=8$. In these examples $z$ is the propagation direction of the optical fields, and both the optical fields and the spin system are initially polarized along the $x$-direction, so that  $\mathbf{S}(0)=S_x$ and $\mathbf{J}(0)=J_x$. 

The most significant feature of these figures is that in all cases $\langle S_y \rangle$ and $\langle S_z \rangle$ remain approximately equal to zero for all times, that is, ${\bf S}(t)$ remains for all practical purposes along the $x$-axis. The present model predicts that $\langle S_y\rangle$ and $\langle S_z \rangle$ undergo small oscillations (Fig.~\ref{fig3}a) whose frequency increases with $J$. This is in contrast to the situation for the Hamiltonian~(\ref{HKU}), where these expectation values remain exactly zero for all times. It is worth noting that the deviation of $\langle S_{y,z}\rangle$ from zero does not change significantly as $S$ is increased,  see Figs.~\ref{fig2} and \ref{fig4}.

For the Hamiltonian (\ref{HKU}) we have that~\cite{ueda-pra-1993}
\begin{equation}
\langle S_x\rangle = S \cos^{2S-1}(\alpha t),
\end{equation}
indicative of large time intervals during which $\langle S_x\rangle = 0$ for large enough values of ${\bf S}$. A similar behavior is found numerically for the Hamiltonian~(\ref{eq:hamiltonian}), as illustrated in Fig.~\ref{fig4}. This is a further indication of the close resemblance between the spin dynamics in the two systems. 

\section{Squeezing}

\begin{figure}
\includegraphics[width=3.6in]{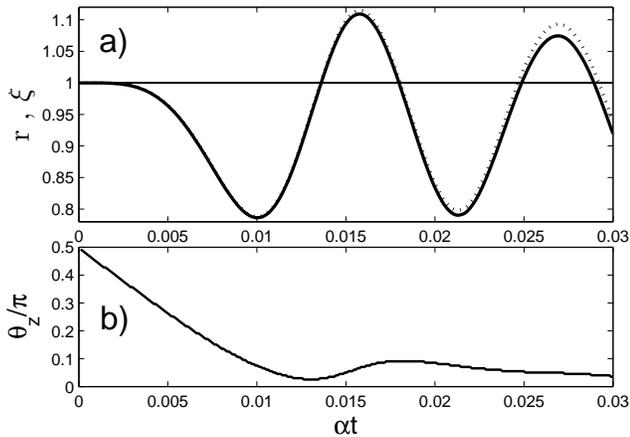}
\caption{(a) Time dependence of $r(t)$ (solid line)  and of $\xi(t)$ (dotted line), with $\xi<1$ corresponding to quantum entanglement \cite{sorensen-nature-2001}.  (b) Direction of spin squeezing in the $(y,z)$ plane as a function of time. The angle $\theta_z$ is defined with respect to $z$. Here $S=80$, $J=120$, and time is in units of $1/\alpha$.}
\label{fig5}
\end{figure}

We now turn to a discussion of spin squeezing. Since to an excellent degree of approximation the spin ${\bf S}$ of the atomic ensemble always points along $x$ we concentrate on the onset of squeezing in the $(y,z)$-plane. We characterize the amount of squeezing and the associated spin-spin entanglement in terms of the ratio
\begin{equation}
r=\Delta S_{\bar{z}}/\Delta S_{\rm coh}
\end{equation} 
between the uncertainty of the spin vector component $S_{\bar{z}}$ and the Standard Quantum Limit  
$$
\Delta S_{\rm coh}=\left(|\langle \mathbf{S} \rangle|/2\right)^{1/2},
$$
that holds for a spin coherent state, as well as via the parameter~\cite{sorensen-nature-2001}
\begin{equation}
\xi^2 =\frac{2S(\Delta S_{\bar{z}})^2}{\langle S_x\rangle^2 + \langle S_{\bar{y}}\rangle^2},
\end{equation}
where
\begin{equation}
\bar{z}=\cos(\theta_z)\hat{z}+\sin(\theta_z)\hat{y}
\end{equation}
is the squeezing direction \cite{ueda-pra-1993}, which corresponds to rotation of the ($y$,$z$)-coordinates about the $x$-axis by an angle $\theta_z$, see Fig.~\ref{fig5}b. As is well known, $\xi^2<1$ is a signature of the inseparability of the density matrix of the $N$-atom system.

\begin{figure}
\includegraphics[width=3.6in]{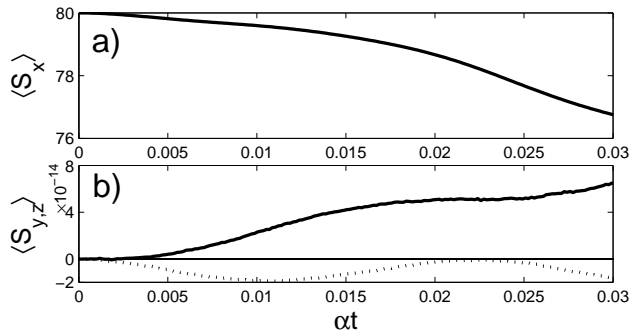}
\caption{Time evolution of the expectation values corresponding to Fig.~\ref{fig5}: (a) $\langle S_x \rangle$; (b) $\langle S_y \rangle$  (solid line) and $\langle S_z \rangle$ (dotted line). Note again the vastly different vertical scales for the two plots. Time in units of $1/\alpha$.}
\label{fig6}
\end{figure}

Figure~\ref{fig5}a shows the time dependence of $r$ (solid line) and $\xi$ (dotted line) as a function of time for $S$ = 80 and $J$=120. As before $z$ is the propagation direction of the optical fields, and both the optical fields and the spin system are initially polarized along  the $x$-direction, so that  $\mathbf{S}=S_x$ and $\mathbf{J}=J_x$.  We note that the dynamics of $r(t)$ and of $\xi(t)$ are perfectly synchronized. This is of course not surprising, since atom-atom correlations are a prerequisite to spin squeezing \cite{bigelow-nature-2001} -- otherwise single-particle uncertainties would simply add up.   Since the spin direction remains essentially along $x$, spin squeezing occurs in the $(y,z)$ plane. It is initially along the $y$-axis, the direction perpendicular to the direction of propagation of the optical field, see Fig.~\ref{fig5}b, and it continuously evolves to along the $z$-axis. In Fig.~\ref{fig6}, we depict the evolution of the spin expectation values $\langle S_x\rangle$ and $\langle S_{y,z}\rangle$, corresponding to Fig.~\ref{fig5} in order to justify that mean spin stays along the $x$-axis.

\begin{figure}
\includegraphics[width=3.6in]{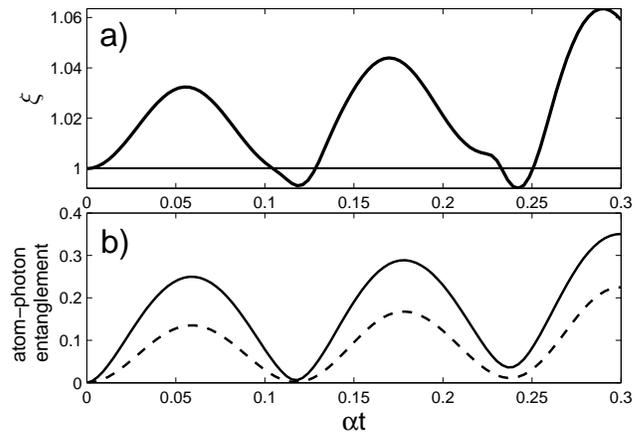}
\caption{Swap of atom-photon entanglement into atom-atom entanglement. (a) Atom-atom entanglement parameter $\xi(t)$ as a function of time for $S=2$ and $J=25$. (b) von Neumann entropy $S_f(t)$ (solid line) of the optical field and Schmidt number $K(t) -1$ (dashed line) for the same parameters. Time in units of $1/\alpha$.}
\label{fig7}
\end{figure}

The interplay between entanglement and spin squeezing is illustrated in Fig.~\ref{fig7}, which shows the reduced von Neumann entropy of the light field
\begin{equation}
S_J=-{\rm Tr} \left ({\hat \rho}_J \log {\hat \rho_J} \right ),
\end{equation}
and the Schmidt number $K$, obtained from the Schmidt decomposition of the state of the atom-field system as
\begin{equation}
|\psi ^{(S,J)}\rangle = \sum_i \sqrt{\lambda_i} |\phi_{i,S}\rangle |\phi_{i,J}\rangle
\end{equation}
as
\begin{equation}
K=\frac{1}{\sum_i \lambda_i^2} = \frac{1}{{\rm Tr}({\hat \rho}_J^2)}.
\end{equation}
Figure~\ref{fig7}b shows that $K$ initially increases, a feature indicative of increased entanglement between the optical field and the atoms. For longer times, though, both $K$ and $\xi(t)$ decrease, with $\xi(t)$ eventually reaching a value below unity indicative of spin squeezing. This behavior is an unambiguous indication of the swapping of entanglement from the atom-field system to atom-atom entanglement.  A similar entanglement swapping mechanism was exploited in the proposal of Takeuchi \textit{et al.} \cite{takeuchi-prl-2005}, a key difference being that in their case a polarization rotator and a mirror were used to achieve entanglement swapping in a two-step process. In contrast,  with the Hamiltonian (\ref{eq:hamiltonian}) the swap process occurs in a single path, without the need for any optical component or cavity. 

\begin{figure}
\includegraphics[width=3.6in]{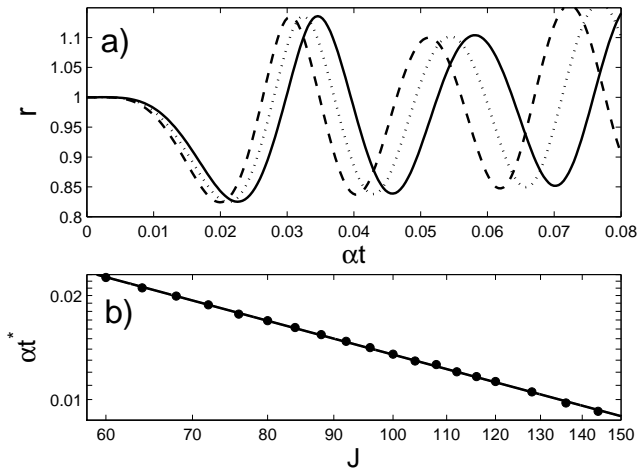}
\caption{(a) Dynamics of the squeezing ratio $r(t)$ for  $J=60$ (solid line), $J=64$ (dotted line), and $J=68$ (dashed line) and fixed ratio $J/S =2$. The time $t^*$ is the time of maximum squeezing.  (b) Log-log plot of $t^*$ versus $J$, with a slope very close to -1.  Time in units of $1/\alpha$.}
\label{fig8}
\end{figure}

Figures~\ref{fig8}a,b show the time $t^*$ at which the first minimum of the squeezing parameter $r(t)$ is reached as a function of the photon number $J$. This dependence is approximately inversely proportional to $J$, as evidenced by the slope of the log-log plot of Fig.~\ref{fig8}b. This indicates that a key parameter in the description of the system dynamics is the scaled dimensionless time $J \alpha t$, and that one can reduce the interaction time $\alpha t$ required to achieve maximum squeezing -- and thereby reduce decoherence effects -- by simply increasing $J$. 

\begin{figure}
\includegraphics[width=3.6in]{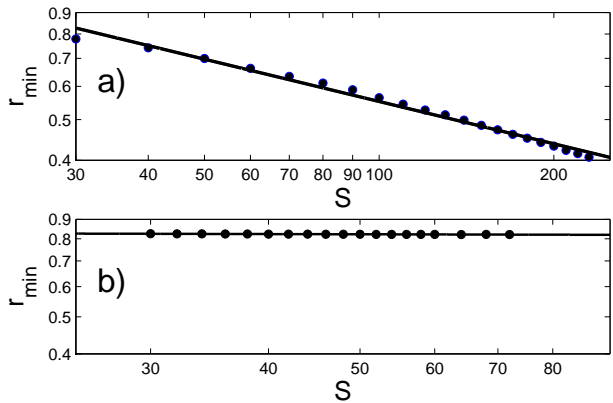}
\caption{Dependence of the minimum squeezing parameter $r(t^*)$ (maximum squeezing) on $S$ for (a) $J=40$ and (b) a fixed ratio $J/S=2$. Both plots are log-log.} 
\label{fig9}
\end{figure}

Figure~\ref{fig9} shows the maximum attainable squeezing (minimum squeezing parameter $r$) as a function of the number of spins $S$ for as fixed photon number $J$ and a fixed ratio of spin to photon number $S/J$. We find that for fixed $J$, $r$ scales approximately as $S^{-1/3}$, similarly to the situation reported in Ref.~\cite{ueda-pra-1993}. However, it is constant for a fixed ratio $S/J$, as would be intuitively expected since in that case the number of photons per atom that can result in entanglement swapping remains constant.

\begin{figure}
\includegraphics[width=3.6in]{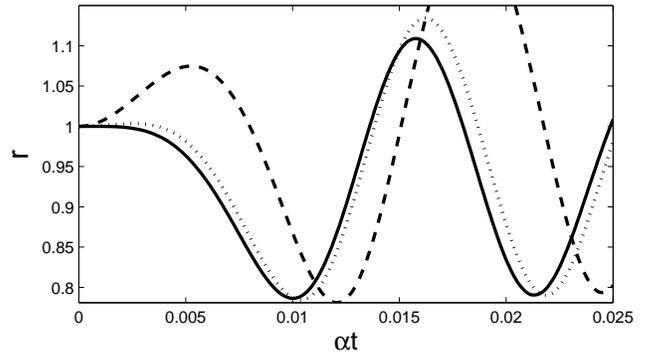}
\caption{Effect of an unwanted diagonal coupling $H_{\rm diag}$ on the squeezing dynamics for  $\beta = 0.1 \alpha$, dotted line, and $\beta = 0.5 \alpha$, dashed line. The idea situation $\beta =0$ is shown as a solid line for comparison. Time in units of $\alpha t$. }
\label{fig10}
\end{figure}

We finally note that in practice, it may not be possible to perfectly eliminate the diagonal terms in the Hamiltonian~~(\ref{eq:hamiltonian}). Figure~\ref{fig10} shows the effects on squeezing of an additional interaction term of the form
\begin{equation}
H_{\rm diag}=\beta \hat{J}_z\hat{S}_z
\end{equation}
illustrating the fact that for moderate coupling strengths of order $\beta = 0.1 \alpha$ it does not substantially effect the squeezing dynamics. Even a much stronger diagonal term, with coupling constant $\beta=0.5\alpha$  does not have a significant impact: it merely shifts the maximum squeezing to later times, after a brief period of anti-squeezing before reaching the first squeezing region. 

\section{conclusion}

In conclusion, we have proposed a simple scheme that permits the realization of spin squeezing via the continuous swapping of atom-photon entanglement into atom-atom entanglement. This scheme can be realized experimentally in alkali atoms driven by two mutually coherent optical fields of orthogonal polarizations. We have also numerically identified scaling laws that permit to predict the time at which maximum squeezing is reached as a function of the number of photons driving the atomic sample, and the maximum achievable squeezing as a function of the number of atoms. This work can be thought of as an extension of the proposal by Takeuchi {\it et al.} where entanglement swapping occurs in a single step rather than their two-step process. The dynamics of the squeezing is a function of the number of photons multiplied by the time variable. 

As a final point we note the complete parallelism between the roles of the atomic spins and the photons, as evidenced by the form of the Hamiltonian~(\ref{eq:hamiltonian}). This indicates that just like the state of the $N$ atoms in not separable, so is the density matrix of the light field, indicative of photon entanglement between the two polarization modes.

Future work will consider the roles of dissipation and decoherence on this spin squeezing mechanism, in particular the role of laser phase and intensity fluctuations. It will also consider the application of spin squeezing in Bose-Einstein condensates magnetically coupled to optomechanical systems in hybrid geometries used for the detection of feeble forces and fields.\cite{steve}

\begin{acknowledgements}
This work was supported by the US Army Research Office, the National Science Foundation, and  the Office of Naval Research. M.E.T. acknowledges support from TUBITAK 2219 fellowship program and from D.P.T. (T.R. Prime Ministry State Planning Organization) National Quantum Cryptology Center.
\end{acknowledgements}


%
%
%

\begin{thebibliography}{99}


\bibitem{mandelwolf} L. Mandel and E. Wolf, {\it Optical Coherence and
Quantum Optics} (Cambridge University Press, Cambridge, 1995).

\bibitem{ueda-pra-1993} M. Kitagawa and M. Ueda, Phys. Rev. A \textbf{47}, 5138 (1993).


\bibitem{sorensen-nature-2001} A. S{\o}rensen, L.-M. Duan, J.I. Cirac, and P. Zoller, Nature \textbf{409}, 63 (2001).

\bibitem{zoller-pra-2003} A. Micheli, D. Jaksch, J.I. Cirac, and P. Zoller, Phys. Rev. A \textbf{67}, 013607 (2003).

\bibitem{esteve-nature-2008} J. Est\'eve, C. Gross, A. Weller, S. Giovanazzi, and M.K. Oberthaler, Nature \textbf{455}, 1216 (2008).

\bibitem{duan-prl-2000-bec} L.-M. Duan, A. S{\o}rensen, J. I. Cirac, and P. Zoller, Phys. Rev. Lett. \textbf{85}, 3991 (2000).


\bibitem{kuzmich-prl-1997} A. Kuzmich, Klaus M{\o}lmer, and E.S. Polzik, Phys. Rev. Lett. \textbf{79}, 4782 (1997).

\bibitem{wineland-pra-1992} D.J. Wineland, J.J. Bollinger, W.M. Itano, F.L. Moore, and D.J. Heinzen, Phys. Rev. A \textbf{46} R6797 (1992).

\bibitem{grom-1995} G.F. Grom and A.M. Kuz'mich, JETP Lett. \textbf{61}, 900 (1995).

\bibitem{hald-prl-1999} J. Hald, J.L. S{\o}rensen, C. Schori, and E.S. Polzik, Phys. Rev. Lett. \textbf{83}, 1319 (1999).

\bibitem{takahasi-pra-1999} Y. Takahashi, K. Honda, N. Tanaka, K. Toyoda, K. Ishikawa, and T. Yabuzaki, Phys. Rev. A \textbf{60}, 4974 (1999).

\bibitem{kuzmich-epl-1998} A. Kuzmich, N.P. Bigelow, and L. Mandel, Europhys. Lett. \textbf{42}, 481 (1998).

\bibitem{kuzmich-prl-2000}  A. Kuzmich, L. Mandel, and N.P. Bigelow, Phys. Rev. Lett. \textbf{85}, 1594 (2000).

\bibitem{vuletic-prl-2010-QND} M.H. Schleier-Smith, I.D. Leroux, and V. Vuleti\'c, Phys. Rev. Lett. \textbf{104}, 073604 (2010).


\bibitem{PS1} Measurement-induced spin squeezing is a powerful tool for initial state preparation, but  it does not provide continuous squeezing as the measurement normally collapses the system to a squeezed state that is not centred about the current value of the mean spin vector.

\bibitem{takeuchi-prl-2005} M. Takeuchi, S. Ichihara, T. Takano, M. Kumakura, T. Yabuzaki, and Y. Takahashi, Phys. Rev. Lett. \textbf{94}, 023003 (2005).


\bibitem{ekert-prl-1993} M. \.Zukowski, A. Zeilinger, M.A. Horne, and A.K. Ekert,
Phys. Rev. Lett. \textbf{71}, 4287 (1993).

\bibitem{bennett-prl-1993} C. H. Bennett, G. Brassard, C. Cr´epeau, R. Jozsa, A. Peres, and W. K. Wootters, Phys. Rev. Lett. \textbf{70}, 1895 (1993).

\bibitem{knight-pra-1998} S. Bose, V. Vedral, and P. L. Knight, Phys. Rev. A \textbf{57}, 822 (1998).

\bibitem{zeilinger-prl-1998} J.-W. Pan, D. Bouwmeester, H. Weinfurter, and A. Zeilinger, Phys. Rev. Lett. \textbf{80}, 3891 (1998).

\bibitem{mustecapPRA2009} M.E. Ta\c{s}g\i n, M.\"{O}. Oktel, L. You,
and \"{O}.E. M\"{u}stecapl\i o\~{g}lu, Phys. Rev. A \textbf{79}, 053603 (2009).

\bibitem{vuletic-prl-2010} I.D. Leroux, M.H. Schleier-Smith, and V. Vuleti\'c, Phys. Rev. Lett. \textbf{104}, 073602 (2010).


\bibitem{yurke-pra-1986} B. Yurke, S.L. McCall, and J.R. Klauder, Phys. Rev. A \textbf{33}, 4033 (1986).


\bibitem{duan-prl-2000} L.-M. Duan, J.I. Cirac, P. Zoller, and E.S. Polzik, Phys. Rev. Lett. \textbf{85}, 5643 (2000).

\bibitem{bigelow-nature-2001} N. Bigelow, Nature \textbf{409}, 27 (2001).



\bibitem{PS3} Weak diagonal terms resulting from the coupling to $F'=2$ manifold can be reduced by up-shifting its energy through coupling to the $F'=2$ and $F'=3$ manifolds  with a strong blue-detuned rf field.

\bibitem{PS4} An anisotropic shift of the excited manifold levels \cite{sorensen-nature-2001,meisner-prl-1999} can also be used to eliminate unwanted diagonal couplings.



\bibitem{danielsteckRb87} Daniel A. Steck, \textit{Rubidium 87 D Line Data}, available online at http://steck.us/alkalidata (revision 2.1.4, 23 December 2010). 

\bibitem{meisner-prl-1999} H.-J. Miesner \textit{et al.},  Phys. Rev. Lett. \textbf{82}, 2228 (1999).

\bibitem{steve} S. Steinke, S. Singh, M.E. Ta\c{s}g\i n, P. Meystre, K. Schwab, and M. Vengallatore, in preparation.

\end{thebibliography}
\end{document}